\documentclass[11pt,a4paper]{article}
\usepackage{color,amsmath,amssymb,soul,graphicx} 

\usepackage[comma, square, numbers, sort&compress]{natbib}
\title{Inhibition in Random Neuronal Networks Enhances Response Variability and Disrupts Stimulus Discrimination}
\author{Netta Haroush$^{1,2}$ and Shimon Marom $^{1,2}$}
\date{%
    $^1$ Network Biology Research Laboratory, Electrical Engineering, Technion- Israel Institute of Technology, Haifa
3200003, Israel.\\%
    $^2$Department of Physiology, Biophysics and Systems Biology, Technion - Israel Institute of
Technology, Haifa 32000, Israel.\\
 \small{*Correspondence and requests for materials should be addressed to N.H. (email:
nharoush@princeton.edu)}\\
    \today
}
\begin{document}

\maketitle

\begin{abstract}
\noindent
Inhibition is considered to shape neural activity, and broaden its pattern repertoire. 
In the sensory organs, where the anatomy of neural circuits is highly structured, lateral inhibition sharpens contrast among stimulus properties. The impact of inhibition on stimulus processing and the involvement of lateral inhibition is less clear when activity propagates to the less-structured relay stations. Here we take a synthetic approach to disentangle the impacts of inhibition from that of specialized anatomy on the repertoire of evoked activity patterns, and as a result, the network capacity to uniquely represent different stimuli. To this aim, we blocked inhibition in randomly rewired networks of cortical neurons \textit{in-vitro}, and quantified response variability and stimulus discrimination among stimuli provided at different spatial loci, before and after the blockade. We show that blocking inhibition quenches variability of responses evoked by repeated stimuli through any spatial source; for all tested response features. Despite the sharpening role of inhibition in the highly structured sensory organs, 
 in these random networks we find that blocking inhibition enhances stimulus discrimination between spatial sources of stimulation, when based on response features that emphasize the relation among spike times recorded through different electrodes. We further show that under intact inhibition, responses to a given stimulus are a noisy version of those revealed by blocking inhibition; such that intact inhibition disrupts an otherwise coherent, wave propagation of activity.

\end{abstract}

\newpage


\section*{Introduction}
Inhibition is a key determinant of structural and functional pattern formation in a wide range of biological phenomena \cite{OSTER1988265,cross1993patternPhysRev,ball1999self,meinhardt2000pattern}. In neural systems, inhibition seems to enrich the repertoire of activity patterns in the developing as well as the mature brain \cite{ben2002excitatory,feller1999spontaneous,steriade1998spike,chervin1988periodicity,buzsaki2012mechanisms,Shew2011}, and there is evidence to that effect also at the behavioral level \cite{hikosaka1985saccadeBicInSC,matsumura1991behavioral,jackson2015inhibition}.
Much of what is known of the impacts of inhibition on neural systems comes from analyses of the sensory envelope. There, where the connectivity of inhibitory neurons is relatively stereotypic and where stimulus evoked activity can be meticulously analyzed, accumulating evidence indicate that inhibition is underlying a core trait shared by all modalities: sharpening of stimulus selectivity and contrast sensitivity by means of lateral inhibition, which depends on a unique structural configuration \cite{hartline1956inhibition,barlow1965mechanism,olsenWilson2008lateralRemoval,arevian2008Olfactory,najac2015intraglomerular}. 
Downstream the sensory envelope, where activity travels through less specialized structures, the impacts of inhibition on sensory processing, as well as the involvement of lateral inhibition, are debated \cite{poo2009odor,isaacson2011inhibition,CarandiniKatzner2011gabaa,TREMBLAY2016260}.\\

Here we investigate the impacts of inhibition on stimulus evoked activity, disentangled from the effects enforced by specialized structures. To this aim we record and analyze responses to stimuli before and after partially blocking inhibition in large-scale random networks of cortical neurons \textit{in-vitro}, on an array of extracellular electrodes. While `not-a-brain,' the setup offers an opportunity to study the effects of inhibition under weaker structural constrains, at high temporal resolution, in multiple sites, and with good pharmacological control over the extent of inhibitory activity. Since many possible conductances could alternatively contribute to a given realization of activity pattern \cite{OLEARY2014809,Marder2014}, we chose to refer to inhibition in the wide sense of driving cells away from an action potential, and preventing propagation of activity. Specifically, we study the impact of bath application of Bicuculline (blocking GABA$_A$ receptors and Ca$^{2+}$ dependent K$^{+}$ channels)
 on trial-to-trial variability, and on stimulus discrimination, for electrical stimuli provided from different spatial locations in the electrode array.
 We show that blocking inhibition in random networks \textit{in-vitro} reduces response variability, and thus sharpens the sensitivity of activity to stimulus location, or have no impact over it, depending on the type of response features taken into consideration. These results suggest that inhibition, at least when embedded in a randomly rewired network, acts as a disperser that enhances response variability and reduces the network capacity to discriminate between stimuli arriving at different spatial locations. Albeit, under intact inhibition the network is less sensitive to specific input loci, and the typical delay between activation of distant cells is shorter, thus manifesting ``small world''  qualities (and possibly related computational benefits). We also show that the network exposed by blocking inhibition is not a different one, but rather share information with the network prior to blocker application. Finally, we bring evidence suggesting that such a dispersing effect of inhibition in large random networks is due to a lateral-inhibition-like contribution: disrupting propagation of activity through near neighbors and thus interrupting an otherwise coherent wave propagation of activity.

\noindent
\section*{Results}

\noindent We study the impact of inhibition on the network capacity to reliably represent different stimuli, in the context of less structure. To this aim, we recorded responses to several stimulation sources, in randomly rewired networks, and compared response properties under control conditions with those obtained under a blockade of inhibition (see Methods). Under this setting, differences between conditions could suggest impacts of inhibition that are independent of specialized structures. We investigated two related aspects of the network response: variability in response to repeating input, and separability of responses to different inputs. The different stimuli here were provided at different spatial coordinates in the network, and can be mapped to different stimuli processed by the same cell assembly. Using an extracellular electrode array to record from \textit{in-vitro} cultures provides incomparable temporal resolution of the group of cells wired together, and a good pharmacological control over blocking inhibition.\\

Specifically, we recorded evoked activity in 17 networks; stimuli were delivered from one fixed source (electrode), or randomly alternated between 2--4 stimulation sources (see Methods).  A response picked by 60 electrodes following a single stimulus (Figure 1A) is demonstrating that the kind of activity recorded in these random recurrent networks is synchronized, but carry a complex spatio-temporal structure. 
In order to quantify response variability and separability of different inputs, we extracted  2 classes of response features: rate, and temporal relations between spikes recorded in different electrodes. For rate features, we used \textit{population firing rate} (Figure 1B), summed over all electrodes; and ``\textit{binary words}" \cite{schneidman2006weak} (Figure 1C), which are firing rate vectors picked by single electrodes, binned such that every time window maximally contains a single action potential (thus bordering precise timing). For the temporal relations between spikes  recorded at different electrodes, we used \textit{first spike latencies} (FSL, Figure 1C), composed of the precise time delay from stimulus to the first spike recorded in 8 of the most active electrodes; and \textit{recruitment order} (Figure 1D), a subset of the FSL, specifying only the participation rank of each electrode of the FSL vector. The relation between these standard response features to the raw raster plot of a single response is readily tractable in Figure 1.  For visual simplicity, only 4 of the most active electrodes (highlighted in red on Figure1A) were used in panels C--D. These compressed representations of different aspects of the raw data, will be used throughout the manuscript for all further analyses. See the Methods section for details on response features extraction.\\

The results section that follows is composed of 3 parts: (1) the impact of  blocking inhibition on response variability under repeating stimulus from a single stimulation source; (2) the impact of blocking inhibition on the discrimination between stimuli provided at different spatial coordinates; and  (3) analyses exploring possible mechanisms involved. \\

\subsection*{Variability in response to repeating input from a single stimulation source}
We recorded responses to repeating input from a single stimulation source (n=8 networks) before and after blocking inhibition (4--5$\mu$M Bicuculine), and evaluated the trial-to-trial variability level under each condition. Since high dose of Bicuculline is known to induce seizure-like epileptic activity \cite{steriade1998spike,keren2016long}, we ensured that Bicuculline doses used in this study were low enough to avoid such an impact (see Methods). Under the partial blockade of inhibition employed here, the network response terminates within $\sim$1 second (and often less), in accordance with physiological timescales of \textit{in-vivo} cortical responses under anesthetized conditions, where the inhibitory contribution is reduced ~\cite{haider2013inhibitionDominates}. Furthermore, networks show enhanced responseiveness to stimulation, as previously reported \cite{haroush2015slow}.

 We find that blocking inhibition reduces variability across trials, for all response features tested. We demonstrate the quenched variability in recruitment order (Figure 2A) and binary words from a single electrode (Figure 2C), taken from a single network under blocked inhibition vs. control conditions (Figure 2B and 2D, respectively). As may be expected, FSL is impacted in a similar manner as recruitment order (not shown); and the reduced variability in population rate (not shown) agrees with previously reported data, after blocking inhibition in these networks \cite{haroush2015slow,Baltz2015}.\\

In order to quantify the reduction in variability across our data set, we resorted to cluster analysis (see Methods): for each response feature and under each condition, we constructed a hierarchical tree of responses, based on their pairwise distances (illustrated in Figure 3A). We then systematically changed the distance cutoff-level and counted the resulting number of clusters (see examples of distance cutoffs in Figure 3A). The intuition being that in the extreme case of a short distance cutoff, each observation (i.e. response feature vector) forms a cluster of its own; whereas at the other extreme, a large enough cutoff would encompass responses from all trials in one single cluster. Following the blockade of inhibition, the reduction in variability is manifested by an over all left shift of the cluster content curves, compared with control conditions, indicating a drop of distance-cutoff for the same number of clusters. We demonstrate this for binary words (Figure 3B, n=64 electrodes in 8 networks), as well as for recruitment order (Figure 3D, n=8 networks). This drop of distance-cutoff for the same number of clusters is also apparent for individual data sets, represented by pairs of control--blocked-inhibition curves, for all response features (demonstrated for binary words and recruitment orders in Figure 3C and 3E, respectively).\\

The data of Figures 2--3 were generated by stimulating the networks at a rate of 0.2 per second. In order to assure that the reduction in variability following a blockade of inhibition is not unique to 5 seconds intervals, we documented the effect at different stimulation rates in the same network (3--40 seconds, $n=4$ networks). Within this range, variability was reduced for all stimulation rates and all features (not shown).\\

\subsection*{Discrimination between responses to stimuli provided at different spatial locations}
Given the stereotypic response pattern from a single stimulation source, induced by blocking inhibition, how does blocking inhibition affects the discrimination of stimuli delivered from different spatial sources? 
One may imagine two extreme pictures: (1) blocking inhibition collapses the network response down to a single pattern, irrespective of the stimulation coordinates. Alternatively, (2) blocking inhibition fragments the network, generating a unique response to each input source; a stiff ``lookup table''. 
 We were particularly interested in the more challenging task of stimulus discrimination that is based on  the activity of cells responding to all stimulation loci, rather than the subgroup of selective cells. This context may be mapped to a scenario of different stimuli being processed by the same cell assembly.\\

To investigate the impact of blocking inhibition on discrimination of responses from different spatial sources, we recorded the network response to randomly alternating stimuli sources (n=9 networks, stimulated at 30 sources all together) before and after bath application of Bicuculline (2-8$\mu$M). We then evaluated separability of responses to different sources using supervised classification and unsupervised separation procedures. To address discrimination of stimuli being processed by the same group of cells, we selected the group of electrodes in which responses were reliably detected for all stimuli, across conditions (see Methods);  and preceded to feature extraction as before. Under blocked inhibition, we find that source discrimination is sharpened, when based on the relation between spike times recorded in different electrodes; while remains unchanged (on average) when based on rate features.\\

We used the contrast between responses pooled from pairs of stimulation sources, to quantify their separability in an unsupervised manner 
(contrast here is calculated based on intra-source and inter-source distances, see Methods). The contrast between pairs of stimulation sources is overall increased under blocked inhibition for recruitment order (Figure 4A), and for FSL (not shown), but not for population rate (Figure 4B) or for binary words from single electrodes (not shown). 
We also used supervised classification with support vector machine (SVM, see Methods) to evaluate discrimination of individual responses, from 2 or 4 stimulation sources. Classification accuracy is improved under blocked inhibition for recruitment order (Figure 4C), and FSL (not shown), but not for for population rate (Figure 4D) or for binary words (not shown) from single electrodes. Similar results were obtained when conducting the SVM procedure on responses from 2 stimulation sources (black markers in Figure 4C and 4D) or on responses from 4 stimulation sources (red markers in Figure 4C and 4D). Note that there are some cases of deteriorated discrimination based on recruitment order (FSL), following the blockade of inhibition; we will revisit these cases in the third part of the Result section.\\

Considering the \textit{relation} between spike times detected in different electrodes, the overall picture seems consistent with a scenario where blocking inhibition partitions the network to a stiff ``lookup" table.
However, discrimination that is based on rate features is indifferent to the quenched variability of these features from individual sources. 

\subsection*{Exploring  possible mechanisms for reduced variability and enhanced stimulus discrimination under blocked inhibition}
Since blocking inhibition enhances the neuronal response gain 
~\cite{isaacson2011inhibition, wilson2012division,olsen2012gain}, it is possible that the emergence of a stereotypic response pattern is a result of the increased spiking probability, and thus a finer sampling rate of the ``complete" spike time series.  To test the contribution of possible ``over-sampling"  to reduced variability under blocked inhibition, we simulated a sub-sampled dataset by randomly erasing spikes from the time series recorded in each electrode under {partially} blocked inhibition (such that the average number of spikes for each electrode is equivalent to that under control conditions). We then repeated the clustering procedure for binary words from the simulated data  (similar to Figure 3B).
Under this ``sub-sampling'', the quenching of pattern content relative to control condition was reduced (as expected), but persisted. This persistent effect suggests that the reduced variability is not merely a result of enhanced firing rate.\\ 

It is often suggested that trial to trial variability is contributed by noise \cite{Faisal2008,mcdonnell2011benefits}. Following this line of arguments, the reduction in variability after blocking inhibition suggests that inhibition contributes noise to neural activity. To simulate the possible contribution of noise to response variability under control conditions, we generated a noisy version of the data recorded under blocked inhibition, by adding random jitters of varying width to its spike-time series. We then extracted response features from the simulated data and compared their pairwise-distances with those computed from the data recorded under control conditions. We find that adding a jitter of ca. $\sim$10 ms yields a comparable mean distance between pairs of responses, to that obtained under control conditions; for  all response features tested.\\

If blocking inhibition exposes a ``reproducible" version of the ``noisy" (or at least variable) response under control conditions, the two response versions should share information. To test whether responses to the same stimulus share information across conditions, we constructed a SVM separation model from each condition (control/blocked inhibition), and attempted to classify responses recorded under the reciprocal condition  (blocked inhibition/control, respectively). Since we were using stimulus discrimination to test for shared information, we focused on the cases where reducing the inhibitory activity had a significant impact on discrimination (FSL, recruitment order).
Conducting this analysis on recruitment order or FSL, we find that using the separation model constructed under the reciprocal condition, in most cases, classification accuracy is still well above chance levels, for data recorded under either conditions (Figure 5A--B). These results suggest that spike time relations under both conditions share information. Furthermore, comparing classification accuracy for responses recorded under control conditions using their own separation model against the accuracy obtained by using the reciprocal separation model {(Figure 5A)}, reveals a correlation between the two methods; such that lossless classification is approached for higher classification accuracy. This correlation suggests that under intact inhibition, better classification accuracy is linked with a weaker contribution of the inhibitory network. Together, these results indeed suggest that responses under intact inhibition are a ``noisy" (or at least variable) version of the reliable responses obtained under blocked inhibition.\\

A third possibility, is that partially blocking inhibition induces a wave like propagation of activity, as familiar from applying high dosage of inhibition blockers, which induces seizure like activity ~\cite{steriade1998spike,huang2010spiral,keren2016long}; but also under physiological conditions, such as those observed in developing networks, where inhibition has yet assumed its re-polarizing impact ~\cite{blankenship2010mechanisms}. 
 It is likely that such coherent traveling waves will facilitate the improved discrimination between stimuli processed by \textit{non-selective} cells, as we show here.
Indeed, under blocked inhibition, FSLs represented on the physical electrode layout show wave like patterns, emerging from the stimulation source (Figure 6B); but much less under control conditions (Figure 6A). An induction of wave like propagation should impact activity beyond the first spikes --- it is accompanied by a topological change, where activity is preferably propagating through near neighbors. In order to quantify such an effect across our dataset, we calculated the conditional firing probability (CFP): the probability of electrode $j$ to detect firing at a time interval $\tau$ following any firing detected in electrode $i$ \cite{eytan2004dopamine,Feber2007}. For each pair of recording electrodes, we registered the delay to the peak of the CFP curve (see left inset of Figure 6C), as the typical delay between the firing times of the two electrodes (see CFP in Methods). The expectation being that under wave like propagation, distal neighbors are activated on longer typical delays. We find that the averaged delay increases with the physical distance for both conditions (data pooled from n=8 networks). However, under  blocked inhibition the peak of the CFP if further delayed for distal electrodes (Figure 6C ), even though propagation velocity is increased. This impact is also apparent for individual networks, from the increased correlation between the typical delay and the distance between electrodes (right inset of Figure 6C) under  blocked inhibition. The elongation of the typical delay under blocked inhibition suggests that the network topology is indeed shifted such that propagation is preferably through near neighbors, as expected under wave propagation of activity. These results also suggest that intact inhibition facilitates activation of distal cells.\\

It is possible that under wave propagation responses to adjacent stimulation sources are less separable, since activity may propagate through similar paths. It was already shown that separability of input sources is sharper the more physically distant the sources are \cite{Kermany2010}. Here we report that this correlation nearly doubles for recruitment order upon blocking inhibition (Fig 6D). Inspecting the difference between the contrast scored on both conditions (Figure 6E) demonstrates that cases where discrimination is not improved, and even deteriorates after blocking inhibition, are confined to shorter distances between stimulation sources. These results suggest that the increased length scale of synchronization, induced by blocking inhibition, sharpens the differences between responses to more distal inputs; while the resolution for distinguishing two inputs apart may deteriorate.\\

Overall, the data reported above suggest that inhibition in random networks is acting as a dispersing agent --- rather than a sharpening one --- enhancing trial-to-trial variability and disrupting stimulus discrimination based on the relations between spike times of different cells. On the other hand, intact inhibition enriches the pattern repertoire of responses, enables faster activation of distal cells, and may facilitate discrimination between adjacent inputs. 
Furthermore, it seems that even at the lack of specialized structures, the impact of random inhibition mimics that of lateral inhibition, by impeding propagation through near-by neighbors.

\section*{Discussion}
We investigated the impact of inhibition on the network capacity to reliably represent different stimuli, at the absence of specialized structures. To this aim, we stimulated random networks \textit{in-vitro}, at multiple locations, and quantified the impact of blocking inhibition on response variability and stimulus discrimination.\\

We show that blocking inhibition quenches response variability from any given source and for all response features tested. The reduction of variability accords with previous reports for single neurons under blocked inhibition, for both evoked and spontaneous activity \cite{najac2015intraglomerular,HAUSSER1997665}. At the network level, it has been recently demonstrated that blocking inhibition in these cell cultures induces highly reproducible  population rate-profiles for both spontaneous and evoked activity \cite{haroush2015slow,Baltz2015,keren2016long}. Furthermore, this macroscopic effect is also reflected in a large body of work on neural avalanches, where blocking inhibition transforms heavy-tailed distributions of event size to narrow bimodal ones \cite{Beggs2003,mazzoni2007dynamics,Shew2009,shew2015maximizing}. Our analyses demonstrate that the reduction in variability is also manifested at the microscopic level, in the fine spatio-temporal patterns of neural activity.\\

We show that under blocked inhibition discrimination that is based on the relations between spike times of different cells (i.e. FSL, recruitment order) is sharpened as a function of the distance between stimulation sources; whereas stimulus discrimination that is based on rate features is unaffected. Both of these observations are against the intuition that tunning-down inhibition is compromising the network ability to differentiate among stimuli. This discrepancy in the role taken by inhibition may be related to the target of our analyses: we were interested in how the same neural population is processing different stimuli, and therefor focused on the less studied case of non-selective cells, rather than the more extensively studied case of selective cells ~\cite{sillito1975completeLostSelect,nelson1994orientation,li2008gaba,CarandiniKatzner2011gabaa}. \\

We show that the overall reduced response variability and improved stimulus discrimination under blocked inhibition are accompanied by coherent traveling waves from the stimulus coordinate. The association between enhanced stimulus discrimination and the induction of wave propagation is non-trivial: if, for instance, under blocked inhibition these network would still maintain ``shortcuts" between distal cells, responses to any stimulus coordinate could potentially collapse to the same spatio-temporal pattern. In this sense, intact inhibition is introducing small-world characteristics: (1) the network is less sensitive to input loci (also apparent in Fig 6D--E); and (2) the effective distance between distal cells is shorter under intact inhibition (Figure  6C). Traveling waves of comparable propagation velocity ($\sim$0.5m/s, Figure 6B) are abundant in sensory cortices ~\cite{sato2012traveling,reimer2010fast,muller2018cortical}. However, their role is unknown, and their existence is considered baffling in the light of stimulus selectivity of cortical cells. Our results suggest that traveling waves may engage a complementary representation strategy to that of stimulus selectivity --- carrying information about the stimulus identity within the spatio-temporal activity patterns of \textit{non-selective} cells.\\

The data is consistent with the interpretation that inhibition in large random networks mimics lateral inhibition, at least in the sense of interrupting the propagation of excitation to nearby neighbors. The spontaneous emergence of such effective lateral-inhibition impact in randomly rewired networks could arise from the fast and coherent manner by which inhibitory activity propagates through strong electrical coupling \cite{gibson2005functional,mancilla2007synchronization}. The spatial correlations involved are thus controlled by the extent of inhibition ~\cite{Petersen2001,karnani2016opening,haider2013inhibitionDominates}, a trace of which is seen in our analyses of CFP after blocking inhibition (albeit intermingled in our data with the dimensions of the electrode array used).\\

We further show that responses to the same stimulation source, recorded under blocked inhibition and control conditions, share information (Figure 5A--B), and that stimulus discrimination is more accurate when responses are more similar to those recorded under blocked inhibition (Figure 5B, note the correlation between classification accuracy of both models). These results suggest that responses under control conditions are a ``noisy" (or variable) version of the reproducible responses under blocked inhibition. There are some indications that inhibition may contribute more to increased variability (whether its origin is stochastic or deterministic), for instance, inhibitory synapses are changing more rapidly \cite{villa2016inhibitory}, and inhibition causes decorrelation of activity in network models \cite{king2013inhibitory,tetzlaff2012decorrelation}, whereas blocking inhibition reduces variability \cite{matsumura1991behavioral,HAUSSER1997665,najac2015intraglomerular}. This quality of inhibiting processes is also familiar from other excitable media systems, where``defects" in transmission, give rise to enrichment of the repertoire of local recurrent activity and entailed adaptations, leading to a dispersion effect \cite{kleber2004arrhythmias,cross1993patternPhysRev,kirkton2011biosyntheticExcit}. 
  \\

Overall, our observations imply that at least for random networks, inhibition acts against the network capacity to maintain and retrieve reproducible information. Albeit, intact inhibition provides enhanced activity pattern repertoire, shortens the activation path between distal cells, and may control the resolution for telling adjacent stimuli apart. The variability induced by inhibition may contribute to the exploration capacity of networks, a basic requirement for systems that adapt to unforeseen challenges. From this perspective, the extent of inhibition determines how amenable to change a given network is. It is therefor tempting to consider the E/I ratio as a dynamic variable that may enhance or restrict network plasticity and enable adaptive network states, for different tasks carried by the same cell-assembly ~\cite{fu2014cortical,Letzkus2015,haroush2015slow,jovanic2016competitive}.\\

\section*{Materials and Methods}

\subsection*{Cell preparation.}
Cortical neurons were obtained from newborn rats (Sprague-Dawley) within 24 hours after birth using mechanical and enzymatic procedures described in earlier studies \cite{Marom2002}. Rats were anesthetized by CO$^{2}$ inhalation according to protocols approved by the Technion's ethics committee. All procedures involving cell preparation and animals handling were  performed in accordance with these guidelines and regulations. The neurons were isolated and plated directly onto substrate-integrated multi electrode arrays. They were allowed to develop into functionally and structurally mature networks over a period of 2 weeks and were used in experiments within the period of 2--6 weeks post plating. The plated neurons cover an area of about $380$ mm$^2$, bathed in a medium supplemented with heat-inactivated horse serum ($5\%$), glutamine ($0.5$ mM), glucose ($20$ mM), and gentamycin ($10~\mu$g/ml), and maintained in an atmosphere of $37^o$C, 5\% CO$_2$ and 95\% air in an incubator as well as during the recording phases.  An array of 60 Ti/Au extracellular electrodes, $30~\mu$m in diameter, spaced $500~\mu$m from each other (MultiChannelSystems, Reutlingen, Germany) was used. The insulation layer (silicon nitride) is pretreated with polyethyleneimine (Sigma, $0.01\%$ in $0.1$ M Borate buffer solution).

\subsection*{Electrophysiology \& Pharmacology.} 
A commercial amplifier (MEA-1060-inv-BC, MCS, Reutlingen, Germany) with frequency limits of 150--3,000 Hz and a gain of x1024 was used for obtaining data. Data was digitized using an acquisition board (PD2-MF-64-3M/12H, UEI, Walpole, MA, USA). Each channel, sampled at a frequency of 16 kHz, detects electrical activity that might be originated from several sources (typically 1--3 neurons) as the recording electrodes were surrounded by several cell bodies. We have used a Simulink-based software for on-line control of data collection [see Zrenner et al.~(2010) for details]. Voltage stimulation was applied in the form of a mono-phasic square pulse $200~\mu$sec 800--950 mV through extracellular electrodes using a dedicated stimulus generator (MCS, Reutlingen, Germany).  Action potentials timestamps were detected on-line by threshold crossing of negative voltage. Detection of synchronous events (Network Spikes, NSs) was performed off-line using a previously described algorithm \cite{Eytan2006} based on threshold crossing of the network firing rate (binned to 3 msec). Once a NS was detected within 400 millisecond following a stimulus, action potentials recorded in all the electrodes within 500 ms following the stimulus were extracted. Post-stimulus time histograms were constructed using a 1 ms time bin, and smoothed with a 5 ms moving average; responses were then screened for a maximum amplitude of at least 1.5 spikes/ms. We performed two types of experiments: in the first,repeating stimuli were applied through a single electrode at a constant interval (5-8 seconds), under control conditions and in the presence of the disinhibiting drug Bicuculline, a blocker of fast GABA transmission (4-5$\mu M$, n=8 networks). In the second type of experiments we provided stimuli from several electrodes (2, 3 or 4 different electrodes; altogether 9 networks) in random orders but at constant intervals (4-8 seconds), under control conditions and blocked inhibitin (2-8$\mu M$ Bicuculline). In 2 out of 9 networks used for stimulus classification experiments, the basal responsiveness was low; in  these two cases we have used 1$\mu$M Bicuculline to increase baseline responsiveness. For blocked inhibition conditions, Bicuculline bath application was administrated as folowing: we first added 1 or 2$\mu M$ of Bicuculline, and waited 10 minutes to make sure that the sabstance impact has stabilized, and then shortly tested the responsiveness from all sources. We then decided whether to add more Bicuculline based on 3 indicators for the level of the blockade: response probability, response latency and response duration \cite{haroush2015slow}. In case response latency was still longer then ca. 50ms, and response probability has not increased for all sources, we increased the dose in 2$\mu M$ steps. In all our experiment, responses did not sustained beyond ca. one second.

\subsection*{Data Analysis.} 
Action potentials recorded within 500 ms following each stimulus were used for the extraction of different response features. Generally, a minimum participation in 90\% of stimuli was set as a limit for inclusion of a channel (electrode) in the analysis of all response features; in analyses of population rate, all electrodes were used. \textit{First spike latencies} were calculated from a subset of the 8 most active electrodes. In cases of a failure to participate in a response, a random value was assigned, pooled from a uniform distribution over the interval [10-500]. This choice was taken in order to avoid trivialization of response classification by SVM according to the missing electrodes. Similar results of the agglomerative analysis (see below) are obtained  when these random values were replaced with zeros.\\ 

\textit{Recruitment order} was defined as the rank of the electrodes, ordered according to their first spike latencies; cases of missed response were assigned an averaged rank (for illustration see Fig1B). \textit{Binary words} were extracted for single electrodes from the 250 ms post stimulus using a 2 ms resolution, resulting in binary vectors. \textit{Population rate} was calculated as described above  over the range of 10-500 ms post stimulus (1 ms resolution, smoothed with 5 ms moving average). \\

\textit{Distances} were computed for pairs of response feature vectors with a $1-cos(\alpha)$ metric. Similar results to those reported here were obtained using other metrics (Levenshtein, correlation, or euclidean). Distances were computed for 40 responses from each of the single source experiments (n=8 networks), and 40 responses from each source for experiments with multiple stimulation sources (n=9 networks).\\

\textit{Supervised and unsupervised classification procedures} were used to quantify dispersion of responses from single sources, as well as for evaluating separability of responses from multiple stimulation sources. To evaluate response dispersion for repeated inputs from a single source, a hierarchic tree was constructed for each network (agglomerative clustering procedure, with $1-cos(\alpha)$ metric); on each iteration the most proximal distance value was taken as the distance between pairs of clusters. Cutoffs limiting the maximal distance allowed between two clusters were implemented. To evaluate separability of responses from multiple sources, a \textit{Contrast} measure adapted from Beggs and Plenz (2004) was calculated for pairs of sources (n=40 per source), defined as follows:\\

$Contrast=\frac{D_{out}-D_{in}}{D_{out}+D_{in}} $\\

With $D_{in}$ being the sum of distances between all pairs of responses to a given stimulation source, $D_{out}$ being the sum of distances between all pairs of inter-source responses (responding to 2 different stimulation sources); using the distance metric of $1-cos(\alpha)$. Note that this measure relies on the ground truth of stimulation identity, and thus only evaluats the relative goodness of separration across conditions.\\ 

\textit{Support vector machine} (SVM) with a Gaussian radial basis function (RBF) kernel, as well as a linear kernel, were used to classify responses from 2 or 4 sources (n=40 per source). Classification performances were averaged over 50 repetitions for each classification process (50\% validation set). For SVM of population rate histograms with a Gaussian radial-based kernel, only the 10-70 ms range were used. \\

\textit{Conditional Firing Probabilities (CFP)} \cite{eytan2004dopamine,Feber2007} were computed for all pairs of electrodes, with a minimal activity threshold of 100 spikes per electrode (within 100 ms from stimulus onset). Binary representations of single electrode firing rate were generated with a temporal resolution of 2 ms. The conditioned probability for electrode \textit{i} to fire within a delay of  $t=\tau$, given that electrode \textit{j} fired at t=0, was computed for $\tau=$ [0,100] at 2 ms steps. The resulting profiles of the CFP were then smoothed with a moving average of 3 consecutive values. The first maximum of each profile was extracted along with its index, which is  referred to as the ``typical delay''. CFP profiles with a maximum lower then 0.05 were excluded. The typical delays were then averaged for distances between electrode \textit{i} and \textit{j} over data pooled from 8 networks.\\

\bibliographystyle{unsrt}  
\bibliography{inhibitionNvariance}
\section*{Author Contributions}
N.H. and S.M. conceived the work, N.H. conducted and analyzed the data, N.H. and S.M. discussed results and wrote
the manuscript.
\section*{Competing interests} 
The authors declare no competing interests.

\newpage
\section*{Figures}
\begin{figure}[h!]
\begin{center}
\includegraphics[trim=20 0 0 0,width=12cm]{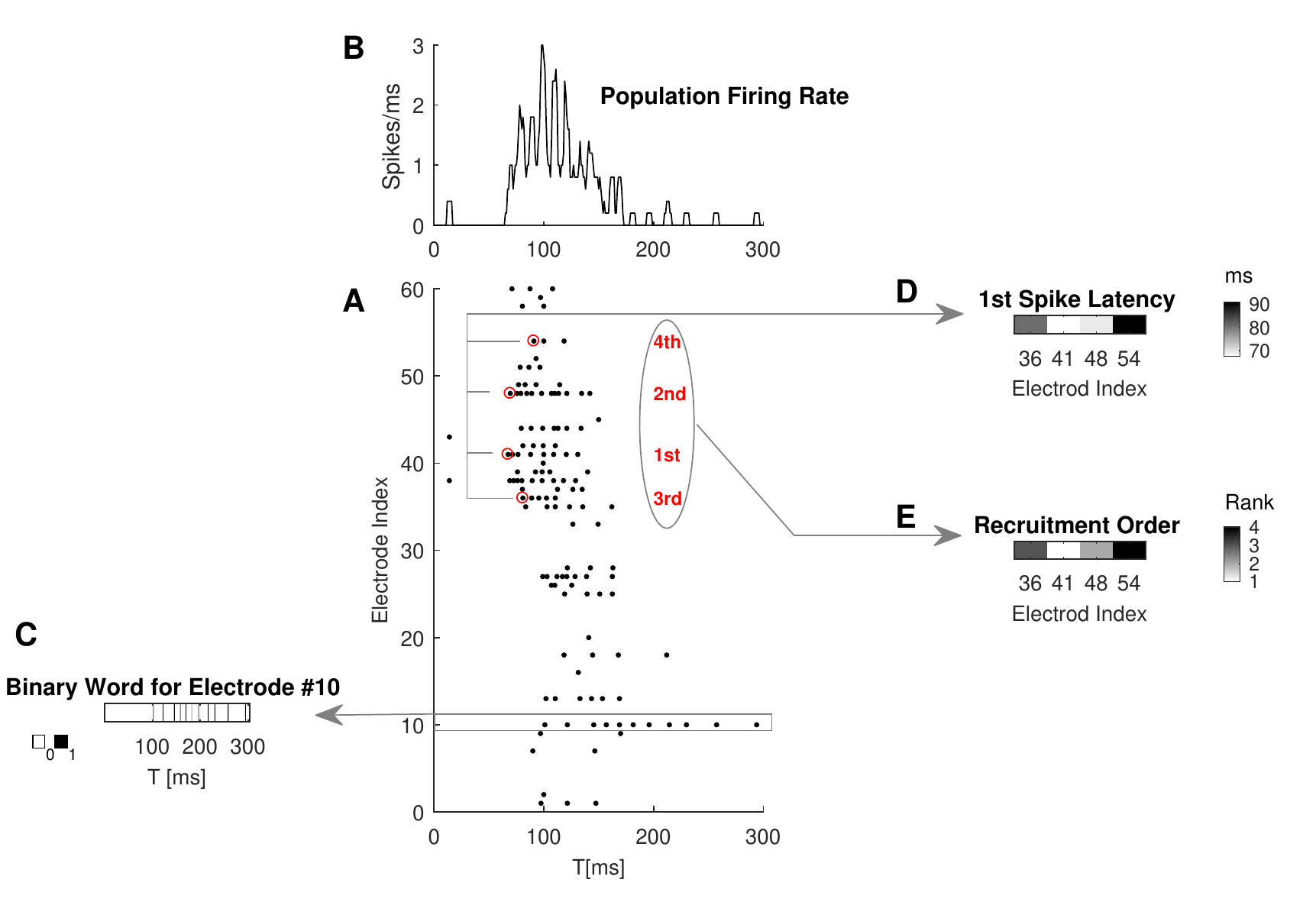}
\end{center}
\caption{\small{{\bf Raw data and feature extraction for further analyses: a response to a single stimulus.} 
A: an instance of a single response, recorded in 60 electrodes during 300 ms post stimulus; the stimulus is applied at t=0 and each dot represent an action potential detected in one of the electrodes. The extraction of response features, subjects to further analyzes in the manuscript, is schematically demonstrated in panels B--E. B: {\bf population rate} is summed over all electrodes; C: Spike time series from single electrodes are converted into {\bf binary words}; D: {\bf first spike latency} is extracted in this example from the first spikes detected in 4 electrodes (highlighted by red circles); and E: {\bf recruitment order} is composed of the participation ranks of the same 4 electrodes (1st, 2nd, etc.). Further details for all feature extraction are available in the Method section.
.}}
\label{fig1}
\end{figure}
 
\begin{figure}[h!]
\begin{center}
\includegraphics[trim=20 0 0 0,width=14cm]{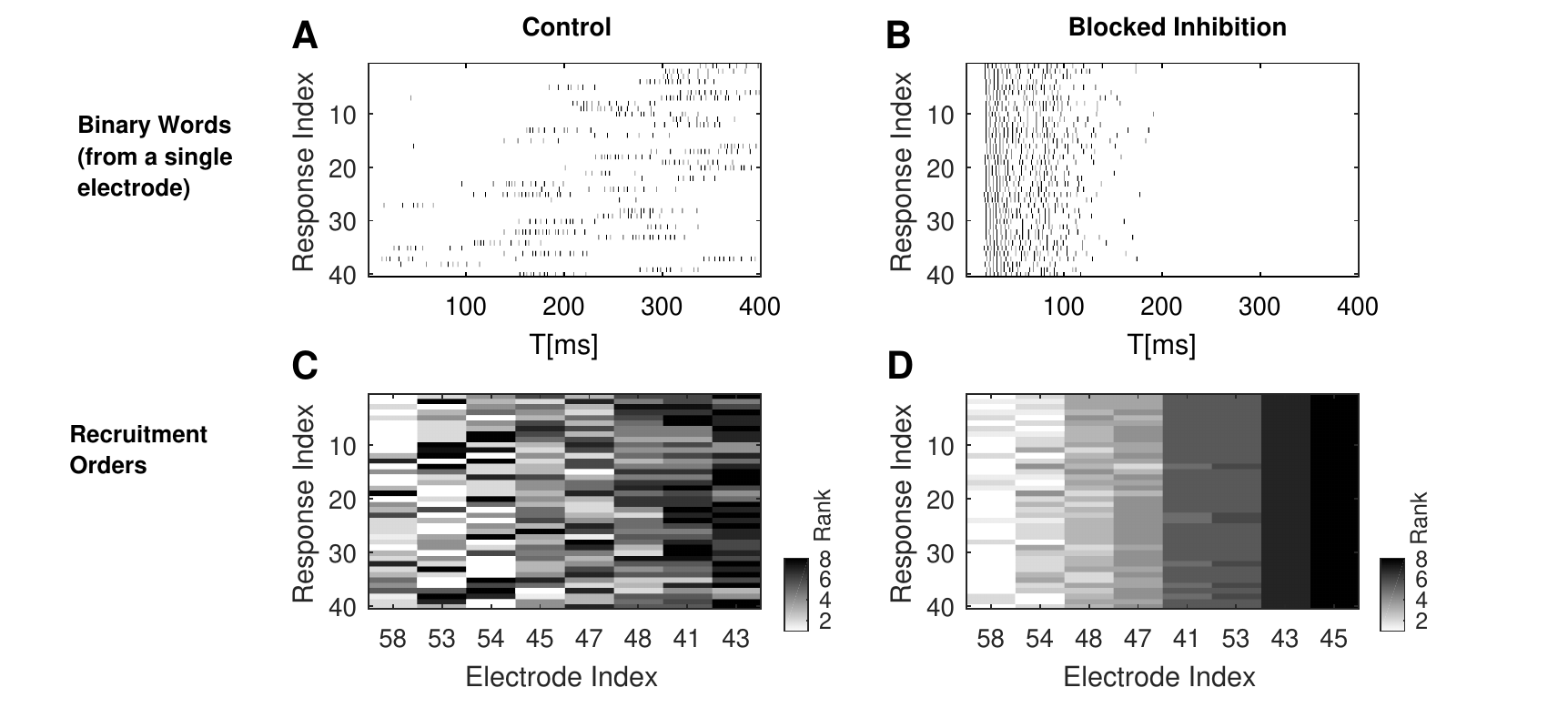}
\end{center}

\caption{\small{{\bf Blocking inhibition reduces trial-to-trial variability in response to a single stimulation source (an example from a single network).}  A: a series of responses picked by a single electrode under control conditions,  represented as {\bf binary words} (each line represents a trial). C: a series of {\bf recruitment order} vectors for 8 of the most active electrodes in the same network, and in response to the same stimuli, under control conditions. Electrodes index is sorted according to their average rank of participation. The impact of blocking inhibition (4-5uM Bicuculline) is shown in panels B\&D: a highly stereotyped pattern is observed across trials, for both B: binary words, and D: recruitment orders.}}
\label{fig2}
\end{figure}

\begin{figure}
\begin{center}
\includegraphics[trim=20 0 0 0,width=14cm]{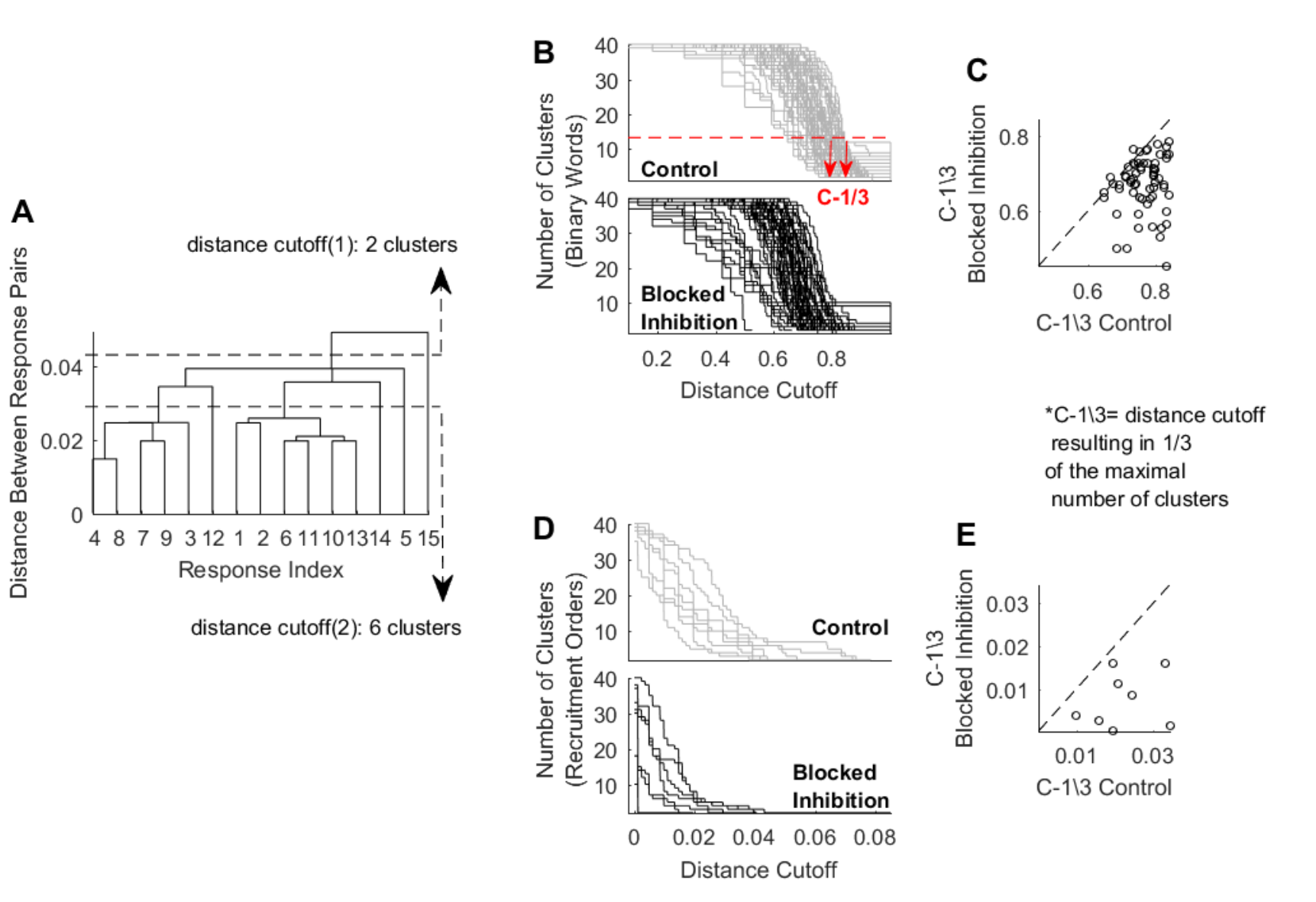}
\end{center}

\caption{\small{{\bf Blocking inhibition quenches the cluster content of response patterns from a single stimulation source.} Variability was quantified for responses to repeating input, using agglomerative clustering (see Methods). For each response feature, distances were computed for all possible pairs of responses in each network (n=8 networks, 40 responses each), and a dendrogram tree was constructed  (for binary words, responses from each electrode were used to construct a dendrogram, n=64 electrodes). A: an instance of such a dendrogram. Cluster content was estimated by systematically changing the distance cutoff along the y-axis, and counting the resulted cluster number. Cutoff(1) and cutoff(2) illustrate the cluster counting procedure, resulting in 2 and 6 clusters, accordingly. B: the number of clusters to which {\bf binary words} are grouped, as a function of the distance cutoff used; under control (top inset) and blocked inhibition (bottom inset) conditions. The cluster content curves show an over all left shift under blocked inhibition, thus the same number of clusters is grouped for lower cutoff values, indicating increased similarity between responses. D: the same for {\bf recruitment orders}. C: quantifies the left shift of pairs of cluster-distance curves obtained before and after the blockade, for binary words. C-1/3: the cutoff value for which the number of clusters drops below one third of the maximal cluster number (illustrated in red arrows on Panel B), is compared between control and blocked inhibition conditions (dashed black line marks unity). Under blocked inhibition, nearly all cluster-distance curves show a marked decrease in C-1/3, as expected for more similar responses. E: the same for recruitment orders.}}
\label{fig3}
\end{figure}

\begin{figure}
\begin{center}
\includegraphics[trim=20 0 0 0,width=10cm]{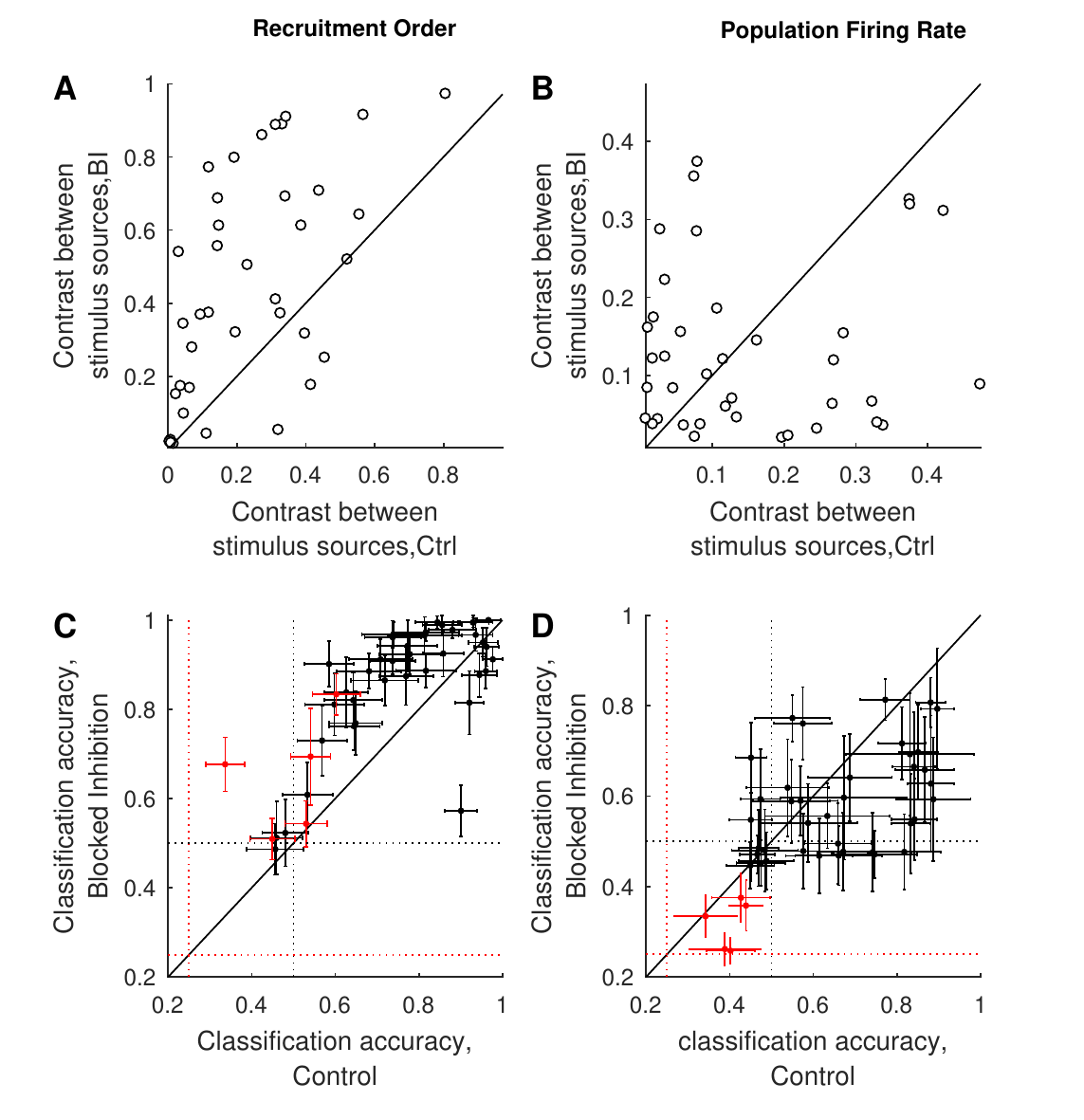}
\end{center}

\caption{\small{{\bf \ Blocking inhibition improves classification accuracy of stimulus locations when based on recruitment orders, but not when based on population firing rates.} We compared stimulus discrimination under blocked inhibition and control conditions. A\&B: show the contrast (see Methods) between pairs of stimulation sources (n=38 pairs from 9 networks), on control (Ctrl) and blocked inhibition (BI) conditions. A: Contrast is enhanced upon blocking inhibition for recruitment order (and for first spike latencies, not shown), but not for population rate (B) or binary words (not shown). Similar results are obtained for supervised clustering (using SVM, see methods), as demonstrated in Panels C--D for recruitment order (C) and population rate (D). Here classification accuracy is presented for pairs of stimulation sources (black, n=38 pairs from 9 networks) and for groups of 4 stimulation sources (red, n=5 networks); error-bars indicate accuracy standard-deviation over repeated classification trials, for which train and test sets are randomized. Dotted lines depict chance levels for pairs of sources (black) and for groups of 4 sources (red). Solid black lines mark unity in all panels.}}
\label{fig4}
\end{figure}

\begin{figure}
\begin{center}
\includegraphics[trim=20 0 0 0,width=9cm]{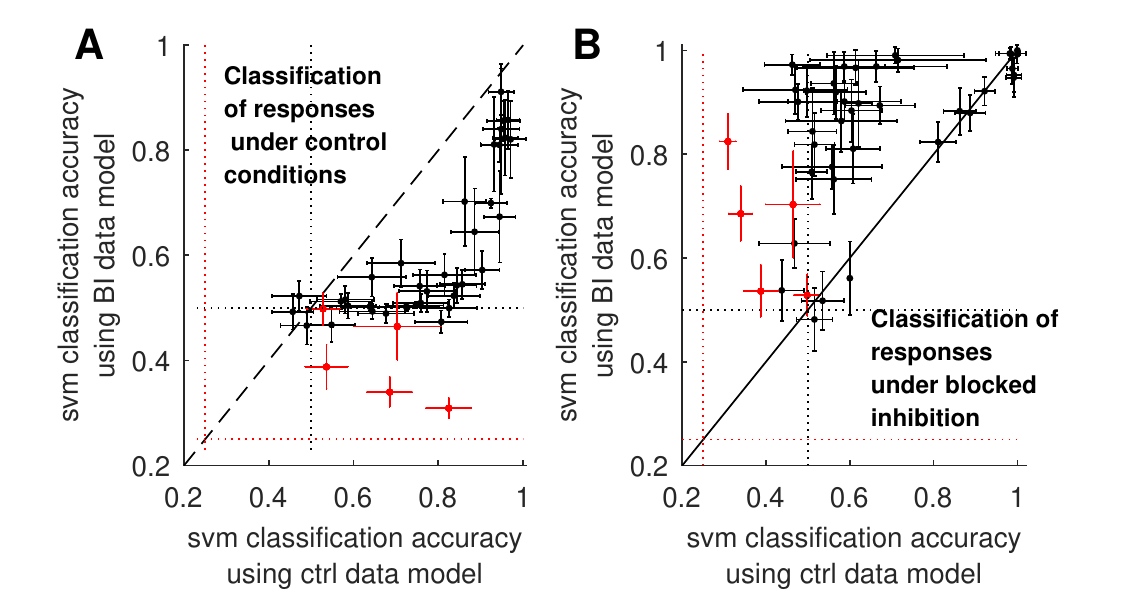}
\end{center}

\caption{{\small{\bf Recruitment order (and first spike latencies) share information across conditions.} To test whether responses under control and blocked inhibition share information, we classified each data set using the reciprocal data training sets. Classification accuracy is shown for recruitment orders of responses recorded under control conditions (A) or under blocked inhibition (B), obtained with training sets recorded under control conditions (x-axis) and under blocked inhibition (y-axis). When using the reciprocal data as training set, classification results based on recruitment order were above chance levels (dotted lines) in most cases (A\& B). This is true for separation between pairs of stimulation sources (black, n=38 sources) as well as for groups of 4 stimulation sources (red, n=5 networks). In panel (A), note the correlation between classification accuracy of both conditions, approaching lossless classification using the blocked inhibition (reciprocal) training set, for the easier tasks under control conditions. Similar results are obtained for first spike latencies (not shown). These results suggest that at least the first spikes recorded in each electrode under control conditions are a noisy version of those recorded under blocked inhibition.}}
\label{fig5}
\end{figure}

\begin{figure}
\begin{center}
\includegraphics[trim=20 0 0 100,width=9cm]{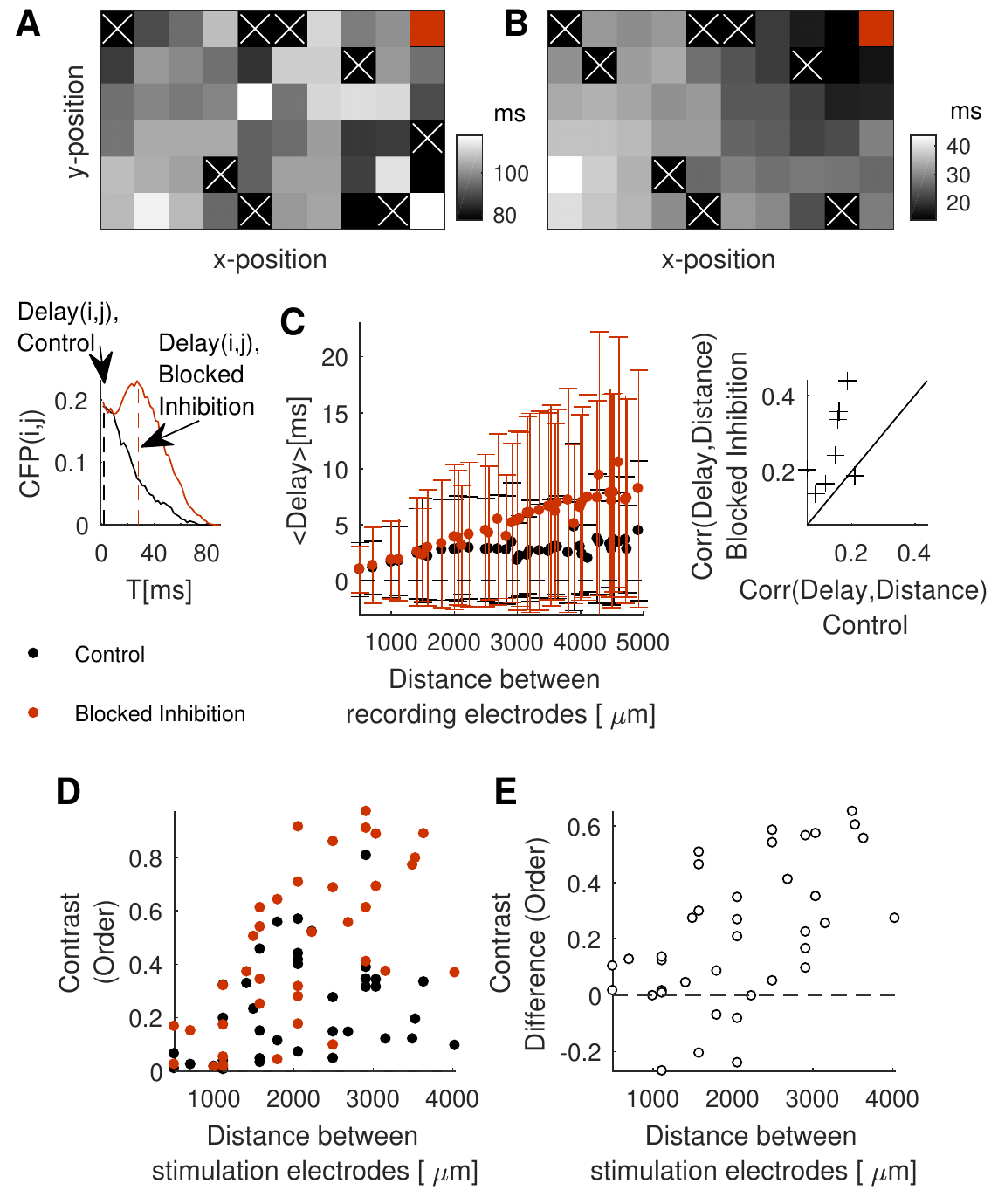}
\end{center}

\caption{\small{{\bf Partially blocking inhibition induces a wave like propagation of activity.} Panels A\&B show instances of first spike latencies, visualized over the x and y positions of the multi-electrode array, from 2 responses to the same stimulation source (stimulus location is marked in red). Isotropic ignition-path is observed under control conditions (A); whereas organized wave of ignition-path is observed under blocked inhibition (B). The color code indicates the first spike latency in milliseconds;  silent electrodes are crossed over.  In order to test whether blocking inhibition induces preferred propagation through near neighbors, we calculated the conditional firing probability (CFP, see text and Methods) between all i,j pairs of electrodes, and tested the correlation of the delay to the CFP$_{i,j}$  peak with the distance between the electrodes i\&j. Instances of the CFP profiles and their typical delay (dashed lines) are shown in Panel C (left inset), under control (black) and blocked inhibition (red) conditions. As expected for traveling waves, under blocked inhibition the mean typical delay is increasing as a function of the distance between the recording electrodes (C, main panel, data is pooled from n=8 networks). This is also apparent in the increase of Pearson correlation between the delay and the distance for individual networks under blocked inhibition vs. control conditions (C, right inset). Note that this is the case even though propagation velocity is increased under blocked inhibition. Traveling waves should also impact the resolution for telling apart adjacent stimulation sources. Panel D shows the contrast between pairs of stimulation sources  as a function of the distance between them, under control (black) and blocked inhibition (red) conditions. Under both conditions stimulus discrimination is improved with the distance, however, this trend is stronger under blocked inhibition. Subtracting the control contrast from the blocked inhibition one (E) exposes that blocking inhibition may worsen discrimination for adjacent stimulation sources, suggesting that inhibition contributes to a finer spatial resolution for telling stimuli apart. 
}}
\label{fig6}
\end{figure}

\end{document}